\documentclass[aps,
article,
superscriptaddress,
twocolumn,
pre,
10pt,
]{revtex4-2}

\usepackage{graphicx}
\usepackage{dcolumn}
\usepackage{bm}

\usepackage[normalem]{ulem}
\usepackage{amsmath,amssymb}
\usepackage{hyperref}

\usepackage{xcolor}

\definecolor{fedeorange}{rgb}{0.98, 0.7, 0.2}

\definecolor{giocolor}{RGB}{0, 150, 100}

\begin{document}

\title{Eco-evolutionary constraints for the endemicity of rapidly evolving viruses}

\author{David Soriano-Pa\~nos}%
\email{sorianopanos@gmail.com}
\affiliation{Departament d’Enginyeria Inform\`atica i Matem\`atiques, Universitat Rovira i Virgili, 43007 Tarragona, Spain.}
\affiliation{GOTHAM lab, Institute for Biocomputation and Physics of Complex Systems (BIFI), University of Zaragoza, 50018 Zaragoza (Spain).}

\date{\today}

\begin{abstract}
Antigenic escape constitutes the main mechanism allowing rapidly evolving viruses to achieve endemicity. Beyond granting immune escape, empirical evidence also suggests that mutations of viruses might increase their inter-host infectiousness. While both mechanisms are well-studied individually, their combined effects on viral endemicity remain to be explored. Here we propose a minimal eco-evolutionary framework to simulate epidemic outbreaks generated by pathogens evolving both their infectiousness and immune escape. Our results reveal that the main driver of viral evolution shifts over time: from intrinsic selection for infectiousness at early stages of the outbreak to antigenic diversification in the transition to the endemic phase. We find that the evolution in both traits during the first epidemic wave plays a critical role in determining long-term viral persistence. Evolution in infectiousness enhances the endemicity of viruses, especially in viruses with lower baseline infectiousness due to the longer duration of their first epidemic wave. Likewise, control policies flattening epidemic curves might increase viral endemicity as a result of the greater antigenic diversity generated in the prolonged epidemic waves. Our results thus prove that the long-term behavior of epidemic trajectories hinges on the complex interplay between both evolutionary pathways and the underlying contagion dynamics.
\end{abstract}


\maketitle


\section*{Introduction}

Modeling the propagation of rapidly evolving viruses poses a major theoretical challenge given the compatibility between the evolutionary and epidemiological time scales~\cite{moya2004population}. Indeed, the evolution of viruses throughout epidemic outbreaks increases the richness of epidemic trajectories and renders complex and fast-changing variants landscapes~\cite{gass2023global,streicker2010host} such as the one recently observed for the SARS-CoV-2 virus~\cite{chen2022global}. Hence, modeling mathematically these intricate dynamics requires moving from the classical compartmental models~\cite{keeling2011modeling} to eco-evolutionary frameworks~\cite{galvani2003epidemiology,makau2022ecological}, explicitly accounting for the rate at which mutations occur, their effects on virus' fitness and the ecological dynamics generated by the competition of the multiple co-circulating strains in the same population. 

Yet eliciting an immune response in their hosts, some rapidly evolving viruses persist in real populations due to the antigenic variation arising from mutations in antibody-binding regions~\cite{carabelli2023sars,smith2004mapping,tuekprakhon2022antibody,van2012evasion}. This evolutionary mechanism facilitates reinfection events by antigenically distant variants that evade pre-existing immunity within the population. The relation between reinfection events and antigenic distances has been incorporated into theoretical models, assuming that variants can be embedded as nodes in genotype networks~\cite{williams2021localization,williams2022immunity,nie2023pathogen} or in discrete~\cite{gog2002dynamics,koelle2009understanding,atienza2023long,yan2019phylodynamic} or continuous~\cite{rouzine2018antigenic,marchi2021antigenic} low-dimensional representations of the antigenic space. Rouzine et al.~\cite{rouzine2018antigenic} show that the interplay between immune presure and virus evolution yields travelling fitness waves in 1D antigenic spaces, sustaining enough viral diversity for the virus to persist in the population. Moreover, antigenic spaces with higher dimensionality can also harbor the speciation of viruses,  splitting at some point into different sublineages which evolve independently~\cite{marchi2021antigenic,yan2019phylodynamic}.

Pathogens' evolution also shapes other non-antigenic traits such as their inter-host transmissibility~\cite{stewart2005empirical,brault2007single} or their virulence~\cite{berngruber2013evolution,walther2004pathogen}, leading to major changes in their associated epidemic trajectories~\cite{zhang2022epidemic,saad2020dynamics,abedon2001bacteriophage} and their impact on public health~\cite{van2021covid,americo2023virulence}.  For instance, recent genomic surveillance data reveals that SARS-CoV-2 virus has undergone both evolution in its infectiousness and antigenic escape~\cite{markov2023evolution,koelle2022changing}. Motivated by this empirical evidence, several theoretical works have proven that the underlying eco-evolutionary dynamics for these viruses intertwine both evolutionary pathways. Namely, non-antigenic traits can evolve towards maximizing either the basic reproduction number ${\mathcal{R}_0}$ or the speed of the antigenic fitness wave at equilibrium as a function of the immune pressure existing in the population~\cite{chardes2023evolutionary,sasaki2022antigenic}.

\begin{figure*}[t!]
\centering
\includegraphics[width=1.7\columnwidth]{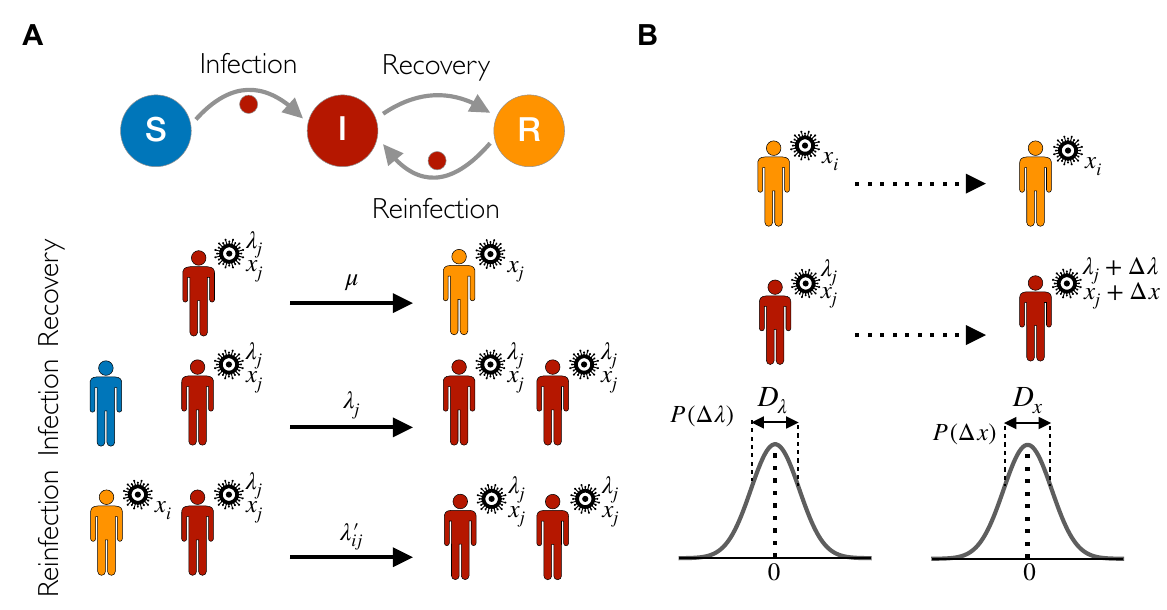}
\caption{{\bf Schematic of the eco-evolutionary model here introduced}. A) Modified Susceptible-Infected-Recovered (SIR) model to account for the reinfection events driven by the immune escape of the virus. Each agent $j$, when infected, recovers at a rate $\mu$, keeping their associated antigenic position $x_j$. Primary infections caused by an infected agent $j$ occur at a rate $\lambda_j$, whereas the contagion rate for a recovered individual $i$ in contact with that infected individual is $\lambda^\prime_{ij}=\Theta(x_j-x_i)\lambda_j \left(1-e^{-(x_j-x_i)}\right)$, where $\Theta(x)$ represents the Heaviside function. Note that the latter increases with the antigenic distance between variants and that reinfection is only possible when $x_j>x_i$, thus assuming that natural immunity against a variant does not wane over time. B) Evolutionary processes of the model. On the one hand, recovered individuals keep their antigenic position. On the other hand, infected individuals evolve both the infectiousness of the virus and their antigenic position, assuming that changes in trait $m$ ($m \in \left\lbrace x,\lambda \right\rbrace$) are drawn from normal distributions, i.e. $\Delta_m \sim \mathcal{N}(0,D^2_m)$, with $D^2_m$ determining its speed of evolution.}
\label{fig:1}
\end{figure*}

As the SARS-CoV-2 virus transitions from an epidemic to an endemic phase~\cite{cohen2022projecting}, there is increasing interest in determining the eco-evolutionary constraints that enable this transition in real populations~\cite{lavine2021immunological,antia2021transition}. To the best of our knowledge, understanding how evolution in both antigenic and non-antigenic traits determines the endemicity of viruses remains an open question. To fill this gap, here we propose a minimal eco-evolutionary framework, extending the classical Susceptible-Infected-Recovered (SIR) model to integrate the evolution of pathogens in both their infectiousness and their immune escape. In absence of the former evolutionary pathway, we show that the probability of a virus becoming endemic depends on its antigenic escape and always increases with the infectiousness of the wild-type variant. In contrast, our stochastic simulations reveal that evolution in non-antigenic traits alters this picture, resulting in a non-monotonic behavior where viruses with low infectiousness display greater endemicity than others with intermediate infectiousness. We derive a heuristic equation capturing this behavior as a result of the trade-off between the minimum infectiousness needed to sustain an endemic scenario and the characteristic evolutionary time scale to reach such infectiousness. To control this evolutionary time scale, we implement control policies flattening epidemic curves and report that extending the first epidemic wave leads to an increase in viral endemicity. To round off the manuscript, we also show  discuss the practical implications of our findings.

\section*{Results}
\subsection*{Model for the coevolution of infectiousness and antigenic escape}
\label{sec:model}
Fig.~\ref{fig:1} sketches the main features of the minimal eco-evolutionary framework here introduced. We assume a constant and well-mixed population of $N$ individuals, where each individual draws $k$ randomly chosen contacts each time step. Our framework is a modified version of the SIR model to harbor reinfection events for those variants escaping the immune response mounted in the population. In this sense, each individual $j$ is characterized by its position in the antigenic space $x_j$, denoting the immune response developed against the virus. For the sake of simplicity, we consider a 1D antigenic space, aligning with previous theoretical frameworks~\cite{sasaki2022antigenic,rouzine2018antigenic}. In addition, we assume that if the individual $j$ is infected, they carry a single variant of the virus, characterized by its infectiousness $\lambda_j$.

Fig.~\ref{fig:1}A illustrates all the processes changing the epidemiological state of individuals in our model. Namely, each infected individual $j$ recovers at a rate $\mu$, preserving the antigenic position $x_j$ of the variant for which they developed their immune response. Regarding contagions, any susceptible individual in contact with an infected individual $j$ contracts the pathogen at a rate $\lambda_j$. Conversely, the rate of reinfection events $\lambda^\prime_{ij}$ is different for each recovered individual $i$, as it should account for the distance between its associated antigenic position and that of the infected agent. 

Following~\cite{gog2002dynamics,rouzine2018antigenic,chardes2023evolutionary}, for each pair $(i,j)$ formed by a recovered individual $i$ interacting with an infected agent $j$, we assume that the contagion rate is given by:
\begin{equation}
    \lambda^\prime_{ij}=
    \begin{cases}
        \lambda_j \left(1-e^{-(x_j-x_i)}\right)  & \text{if } x_j \geq x_i \\
        0 & \text{otherwise}
    \end{cases}
\end{equation} 

Consequently, reinfection events only occur when $x_j>x_i$. This assumption is needed to avoid reinfections of variants already contracted in the past and account for the immune memory host. For all contagions, the agent contracting the pathogen inherits the variant of the infected individual. 

Fig.~\ref{fig:1}B shows the evolutionary processes changing the epidemiological parameters of the circulating viral strains. We assume that mutations just occur in infected individuals. To reflect the impact of these mutations, for each time step, we update the antigenic position of each infected individual $j$, $x_j$, by a factor $\Delta x$. Inspired by previous works~\cite{zhang2022epidemic,rouzine2018antigenic,chardes2023evolutionary}, we assume that changes in antigenic position are drawn from a zero-mean Gaussian distribution whose variance $D^2_x$ determines the speed of evolution. Regarding recovered individuals, we assume their antigenic position to remain constant over time until they get reinfected by another circulating strain. Note that negative changes in antigenic position, yet occurring in mutations inside hosts, are not selected at the population level and the overall behavior corresponds to a travelling wave of strains moving at a positive speed in the 1D antigenic space~\cite{rouzine2018antigenic}. Moreover, we assume that changes in infectiousness are drawn from another uncorrelated zero-mean Gaussian distribution whose variance is denoted by $D^2_\lambda$. A complete description of the stochastic simulations performed to obtain the epidemic trajectories shown in the manuscript is given in Appendix \hyperref[sec:appendixA]{A}.

\subsection*{Endemicity of viruses without evolution in infectiousness}
\begin{figure}[t!]
\centering
\includegraphics[width=1.0\columnwidth]{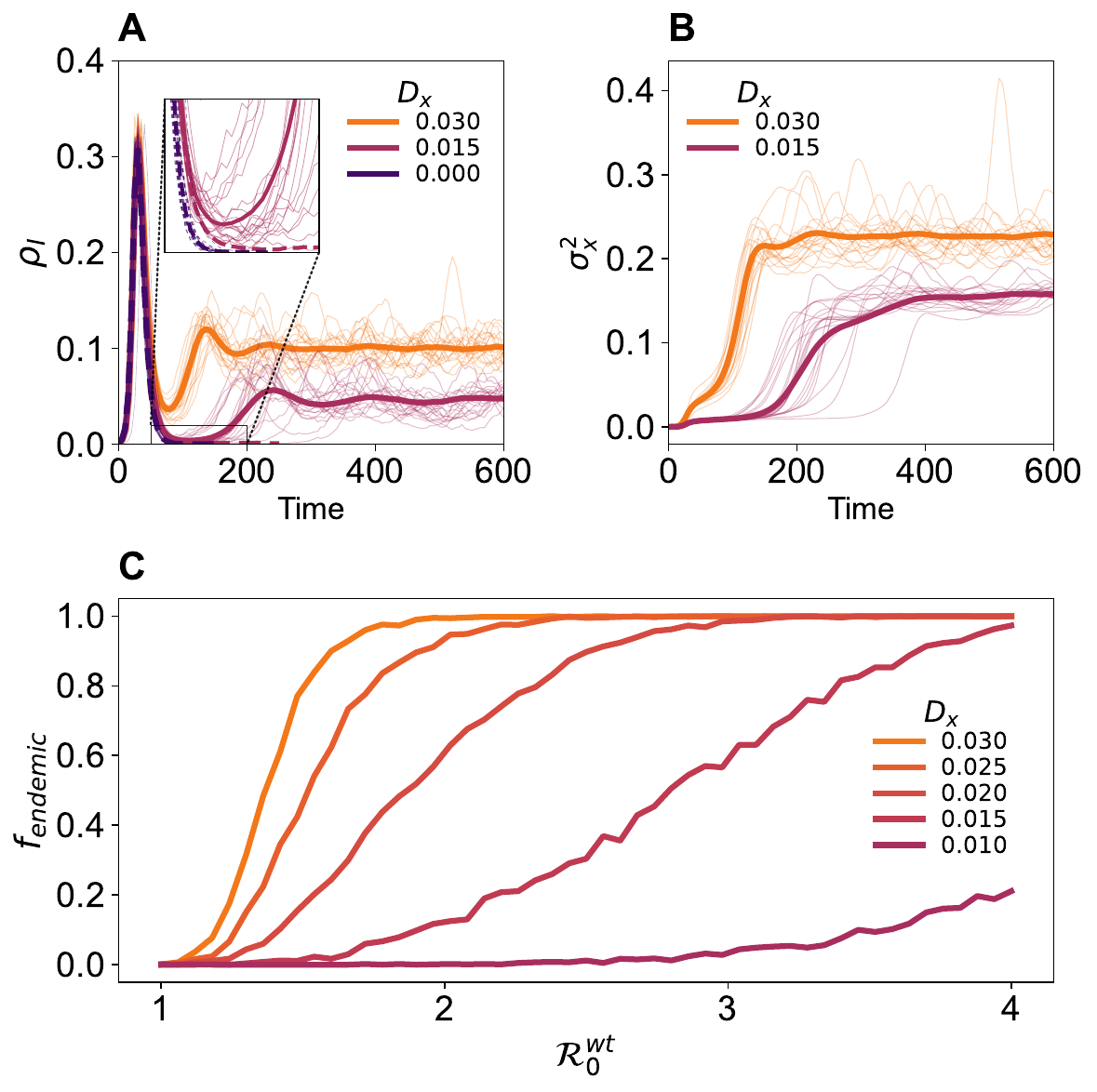}
\caption{{\bf Epidemic trajectories for viruses not evolving their infectiousness}. A) Time dependence of the fraction of population in the infected state $\rho_I$. B): Time evolution of the variance of the distribution of variants in the antigenic space $\sigma_x^2$. In both panels, thick solid (dashed) lines shows the values for those viruses becoming (not becoming) endemic in the population obtained by averaging 200 epidemic outbreaks whereas thin lines represent a sample of 20 individual trajectories in both cases. In these panels, the basic reproduction number of the wild-type variant of the pathogen ${\mathcal{R}_0^{wt}}$ is set to $\mathcal{R}^{wt}_0$=3. C): Fraction of epidemic outbreaks surviving in the population after $t=1000$ days $f_{endemic}$ as a function of ${\mathcal{R}^{wt}_0}$. The results shown in this panel have been obtained by simulating $1000$ epidemic outbreaks for each pair of ($\mathcal{R}^{wt}_0$,$D_x$) values. In all panels, line color represents the value of speed of evolution in antigenic position $D_x$. For the simulations, in all panels we assume $N=10^4$ individuals, $I_0=10$ initially infected agents, and we set the recovery rate to $\mu=1/7$ days$^{-1}$ and the number of daily contacts to $k=10$ interactions.}
\label{fig:2}
\end{figure}

To understand the transition from the epidemic to the endemic phase, let us first neglect the evolution of the virus in infectiousness by setting $D_\lambda=0$ and simulate different epidemic outbreaks varying the speed of evolution in the antigenic space $D_x$. Unless otherwise stated, throughout the manuscript we fix the recovery rate of the disease to $\mu=1.0/7$ days$^{-1}$ and consider epidemic outbreaks spreading across populations of $N=10^4$ individuals, each one making $k=10$ contacts per day. For all the simulations, we assume an initial prevalence of the disease $\rho_I (0) = 0.001$ on average and consider that all the individuals initially infected share the same wild-type strain $j$, characterized by $x_j = 0$ and $\lambda_j = {\cal R}_0^{wt} \mu/k$, where ${\cal R}_0^{wt}$ denotes its basic reproduction number.

Fig.~\ref{fig:2}A depicts the three different epidemic regimes which can be found in the model for a pathogen with $\mathcal{R}^{wt}_0=3$. Recall that the basic reproduction number quantifies the number of contagions triggered by one infected individual in a healthy population. Therefore, when $\mathcal{R}^{wt}_0>1$, the pathogen always produces a first epidemic wave in the population. Nonetheless, the long-term behavior of the epidemic trajectories does depend on the immune escape mechanism, parametrized with $D_x$. In absence of immune escape, i.e. $D_x=0$, the model reduces to the standard SIR model characterized by a single epidemic wave followed by the extinction of the infected population. Conversely, for $D_x=0.03$, a damped oscillatory behavior towards an endemic equilibrium appears for all epidemic trajectories. In between these two extremes ($D_x=0.015$), richer epidemic dynamics are generated by the model, as the first epidemic wave is typically followed by long periods of very low epidemic incidence. Such periods act as epidemic bottlenecks, leading to the extinction of some outbreaks due to stochastic fluctuations. Others, however, are able to persist in the population as a result of the enlargement of the distribution of variants in the antigenic space, as shown in Fig.~\ref{fig:2}B. Comparing the endemic realizations, higher $D_x$ values lead to endemic scenarios with a more acute disease prevalence, as reported in~\cite{rouzine2018antigenic}. 

Fig.~\ref{fig:2}C shows the fraction of epidemic outbreaks surviving in the population after $t=1000$ days, denoted in what follows by $f_{endemic}$, as a function of both ${\mathcal{R}}_0^{wt}$ and $D_x$. There, the three aforementioned epidemic regimes become more evident. Moreover, virus endemicity increases monotonically with the basic reproduction number of the wild-type variant regardless of the speed of evolution in the antigenic space $D_x$. Likewise, for a given $\mathcal{R}_0^{wt}$ value, accelerating the evolution in the antigenic space, i.e. increasing $D_x$, also enhances virus endemicity, making immune escape more likely.

\subsection*{Endemicity of viruses with evolution in infectiousness}

So far, we have neglected the evolution of virus infectiousness, retrieving the three dynamical regimes obtained in the literature for the SIR model with immune escape and the expected monotonic behavior of endemicity with virus infectiousness. Hereafter, we include such evolutionary pathway and address how the joint evolution of both antigenic and non-antigenic traits alters this picture. We first fix the evolution speed of the different traits by setting $D_x = 0.015$ and $D_\lambda = 3\cdot 10^{-4}$ and study the epidemic trajectories for three different pathogens differing in the basic reproduction number of their wild type variants ${\cal R}_0^{wt}$.  

\begin{figure}[t!]
\centering
\includegraphics[width=1\columnwidth]{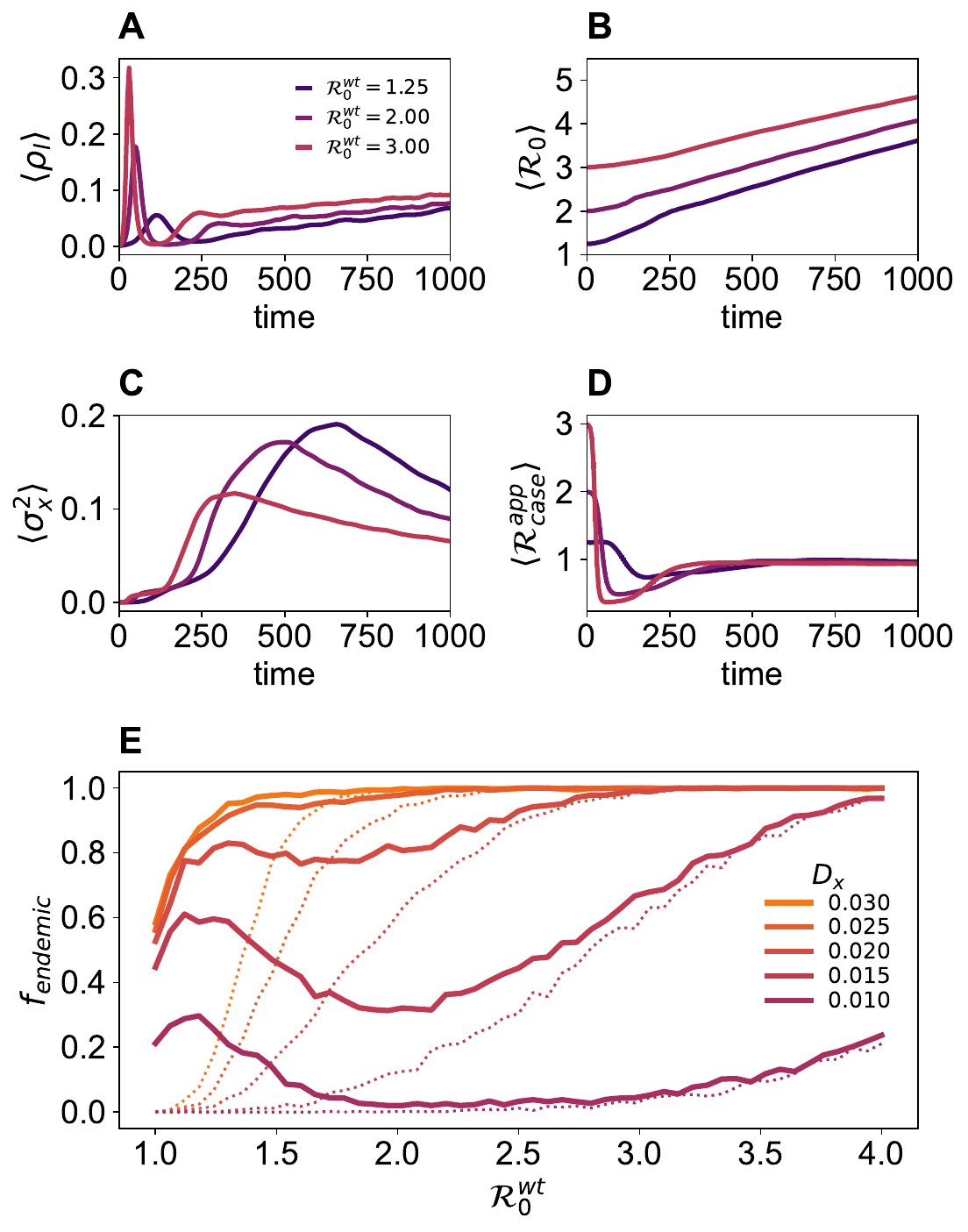}
\caption{{\bf Eco-evolutionary dynamics under evolution of antigenic and non-antigenic traits}. A)-D): Time evolution of different epidemiological quantities and virus traits in endemic epidemic outbreaks. The quantities shown correspond to: (A) fraction of infected population $\rho_I$, (B) basic reproduction number ${\cal R}_0$, (C) variance of the distribution of strains across the antigenic space $\sigma^2_x$ and (D) an estimation of the case reproduction number $R^{app}_{case}$ (see Appendix \hyperref[sec:appendixC]{C}). In all these panels, line color corresponds to the basic reproduction number of the wild-type variant ${\cal R}_0^{wt}$.The symbol $\langle \cdot \rangle$ denotes that each curve is the result of averaging the individual curves of all endemic realizations observed after simulating $1000$ epidemic outbreaks for each ${\cal R}_0$ value. The speeds of evolution in the infectiousness and antigenic spaces are set to $D_\lambda=0.0003$ and $D_x=0.015$ respectively. E): Endemicity $f_{endemic}$ of the virus as a function of the basic reproduction number of the wild-type variant ${\cal R}_0^{wt}$. Line color here represents the speed of evolution in the antigenic space $D_x$. Solid (dotted) lines represent the values found in presence (absence) of evolution in the infectiousness space by setting $D_\lambda=0.0003$ ($D_\lambda=0$). In all the panels, the rest of epidemiological parameters are the same as in Fig.~\ref{fig:2}.}
\label{fig:3}
\end{figure}

Fig.~\ref{fig:3}A shows that evolution in infectiousness allows mildy infectious viruses, e.g. ${\cal R}_0^{wt}=1.25$ or ${\cal R}_0^{wt}=2$, to become endemic in the population. Note that this result comes in stark contrast with those reported in Fig.~\ref{fig:2}C, as no endemic realization is observed for this combination of parameters when just accounting for the evolution in antigenic traits. Regarding the shape of the temporal trajectories, the early stages of the outbreak qualitatively resemble the ones shown in Fig.~\ref{fig:2}A with the presence of a first prominent epidemic wave followed by an epidemic bottleneck. Conversely, the late stages of the outbreak are not characterized by a fixed prevalence of the disease as in the former case. Interestingly, regardless of the infectiousness of the wild-type variant, the epidemic prevalence seems to grow at a (roughly) constant pace on average once the virus has entered the endemic phase. 

Let us now analyze how each virus trait evolves over time. Fig.~\ref{fig:3}B shows the time evolution of the associated basic reproduction number ${\cal R}_0$ for each pathogen here considered. For each realization, we consider the mean infectiousness of the circulating variants at a given time step $t$, $\bar{\lambda} (t)$, to compute this indicator, yielding ${\cal R}_0 (t)= \bar{\lambda} (t)k/\mu$. In all cases, we can observe how infectiousness evolves quicker at early than at late stages of the outbreak. This phenomenon becomes more pronounced when considering viruses with low initial infectiousness and high $D_\lambda$ values (see Appendix \hyperref[sec:appendixB]{B}). Consequently, our results reveal how the underlying eco-evolutionary dynamics yields an accelerated evolution in infectiousness for innocuous pathogens ($\mathcal{R}_0^{wt}=1.25$) compared to that experienced by more infectious pathogens, e.g. those ones with ${\cal R}_0^{wt}=3$.

The evolution of antigenic traits can be characterized by the distribution of strains across the antigenic space.  In contrast with the evolution in infectiousness, we observe in Fig~\ref{fig:3}C how evolution in the antigenic space mainly occurs in the transition to the endemic regime after the epidemic bottleneck, reflected by the sudden enlargement in the variants distribution observed for the three pathogens. Comparing the different curves, we notice the lower the infectiousness, the higher the peak of antigenic diversity reached by the pathogen. This result is intuitive, as those pathogens with less infectiousness should accumulate more antigenic diversity to survive in the population. Note that coupling both evolutionary pathways also has strong implications for the long-term evolution of the antigenic diversity. Specifically, antigenic diversity is not stabilized after the initial increase, as in Fig.~\ref{fig:2}B, but is partially lost as the virus gets more infectious to eventually reach a stable but lower value.

To characterize the emergent relationship between both evolutionary pathways, in Appendix \hyperref[sec:appendixC]{C} we compute a theoretical approximation for the case reproduction number ${\cal R}^{app}_{case}(t)$ representing the expected number of contagions made by one individual infected at a given day $t$~\cite{wallinga2004different}. The time evolution of this quantity (Fig.~\ref{fig:3}D) illustrates how antigenic and non-antigenic traits compensate one another to produce the same steady growth in the epidemic curves at late stages of the outbreak. This confirms that, despite the absence of any biological trade-off, the combination of immune pressure in the antigenic space and evolution in infectiousness yields universal trajectories for the evolution of pathogens in our model. 

We now focus on our primary research question and study the eco-evolutionary constraints for the endemicity of evolving viruses. Interestingly, Fig.~\ref{fig:3}E reveals that evolution in infectiousness drastically changes the picture observed in its absence (Fig.~\ref{fig:2}C). As stated before, evolution in infectiousness promotes viral endemicity, as the fraction of surviving realizations is always higher than in absence of such evolutionary pathway. More strikingly, in situations with little antigenic diversity, i.e. low $D_x$ values, evolution in infectiousness gives rise to a non-monotonic behavior of the virus endemicity with the infectiousness of the wild-type variant. For instance, the curve corresponding to $D_x=0.015$ shows that viruses with an initial reproduction number of $\mathcal{R}_0^{wt}\simeq 1$ are more prone to persist in the population than others with $\mathcal{R}_0^{wt}\simeq 2$. Note that this unexpected behavior disappears when immune pressure loses relevance, i.e. for high $D_x$ values or high $D_\lambda$ values (see Appendix \hyperref[sec:appendixB]{B}), retrieving the monotonic behavior typically reported in the literature. 

For the sake of completeness, we also consider a scenario where viral evolution is bounded by a transmission-recovery trade-off accelerating the recovery rate of hosts for more transmissible pathogens (see Appendix \hyperref[sec:appendixD]{D} for more details). Biologically, the latter assumption is justified as an enhanced immune response for pathogens which replicate more efficiently within the host~\cite{alizon2008transmission}. Our results show that the interplay between evolution and endemicity remains robust as long as the relative increase in the recovery rate is negligible. Nonetheless, when the trade-off gains relevance, the non-monotonic behavior is lost. In this case, the initial evolution in infectiousness shortens the characteristic time scale of outbreaks, thus limiting immune escape. Moreover, under the transmission-recovery trade-off, viruses are able to find evolutionarily stable strategies (ESS), thus eventually slowing down the evolution in infectiousness and therefore the chances for pathogens to become endemic. More future research will be relevant to determine how the phenomena here reported are sensitive to other biological trade-offs.

\subsection*{Evolution of infectiousness in the first epidemic wave explains the non-monotonic endemicity with ${\cal R}_0^{wt}$}
Our results from stochastic simulations reveal that the interplay between immune pressure and evolution in infectiousness favors the endemicity of weakly infectious pathogens. To understand the origin of such phenomenon, we first realize that, in absence of infectiousness evolution (Fig.~\ref{fig:2}C), there is a critical value of the basic reproduction number ${\cal R}^{wt,C}_0 (D_x)$ below which viruses become extinct. This critical value is lower for viruses with efficient immune escape, i.e. high $D_x$ values. Inspecting visually the critical values in the endemicity curves for viruses now evolving their infectiousness (Fig.~\ref{fig:3}E), we notice that the unexpected non-monotonic behavior involves pathogens whose basic reproduction number falls below ${\cal R}^{wt,C}_0 (D_x)$. Conversely, the region ${\cal R}_0^{wt} > {\cal R}^{wt,C}_0 (D_x)$ retrieves the monotonically increasing behavior with the infectiousness of the wild-type variant. From this qualitative analysis, it becomes clear that the non-monotonic behavior previously described arises from the evolutionary dynamics undergone by viruses with ${\cal R}_0^{wt} < {\cal R}^{wt,C}_0 (D_x)$ to reach that critical value before becoming extinct. 

To further support our intuition, we seek to analytically derive the eco-evolutionary constraints determining the endemicity of rapidly evolving viruses. Let us first neglect the evolution in infectiousness. In this case, the state of our system at time $t$ is characterized by the distributions, $\rho_I(x,t)$ ($\rho_R(x,t)$), defined as the fraction of the population in the infected (recovered) state carrying (with immunity to) the strain $x$ at time $t$. Considering antigenic escape and reinfection events, the time evolution of the latter is given by~\cite{rouzine2018antigenic,chardes2023evolutionary}:
\begin{align}
    \frac{d\rho_R(x,t)}{dt} &= \mu \rho_I(x,t) \nonumber \\ &- \lambda k \rho_R(x,t)\int_x^{\infty} \rho_I(x^\prime,t)\left(1-e^{-(x^\prime - x)}\right)dx^{\prime}\;,
    \label{eq:det1}
\end{align}
where the first term corresponds to the recovery of infected individuals and the second term encodes the loss of recovered individuals due to reinfection events by variants with different antigenic features.

Unfortunately, the underlying stochasticity of both evolutionary and epidemiological processes makes the analytical treatment of this model quite cumbersome, yet some theoretical results can be obtained in specific circumstances~\cite{chardes2023evolutionary,rouzine2018antigenic}. To overcome this issue, we move to a more simplified eco-evolutionary framework, where we assume that both infectiousness and antigenic position grows linearly with time following their evolution inside infected hosts. Note that, in doing so, we cannot capture some of our previous findings, such as the accelerated evolution of the most innocuous viruses at early stages of the epidemic outbreak. Nonetheless, this constant evolution, yet unrealistic, captures the increasing infectiousness and antigenic position due to the selection driven by epidemiological processes.

\begin{figure}
\centering
\includegraphics[width=1\columnwidth]{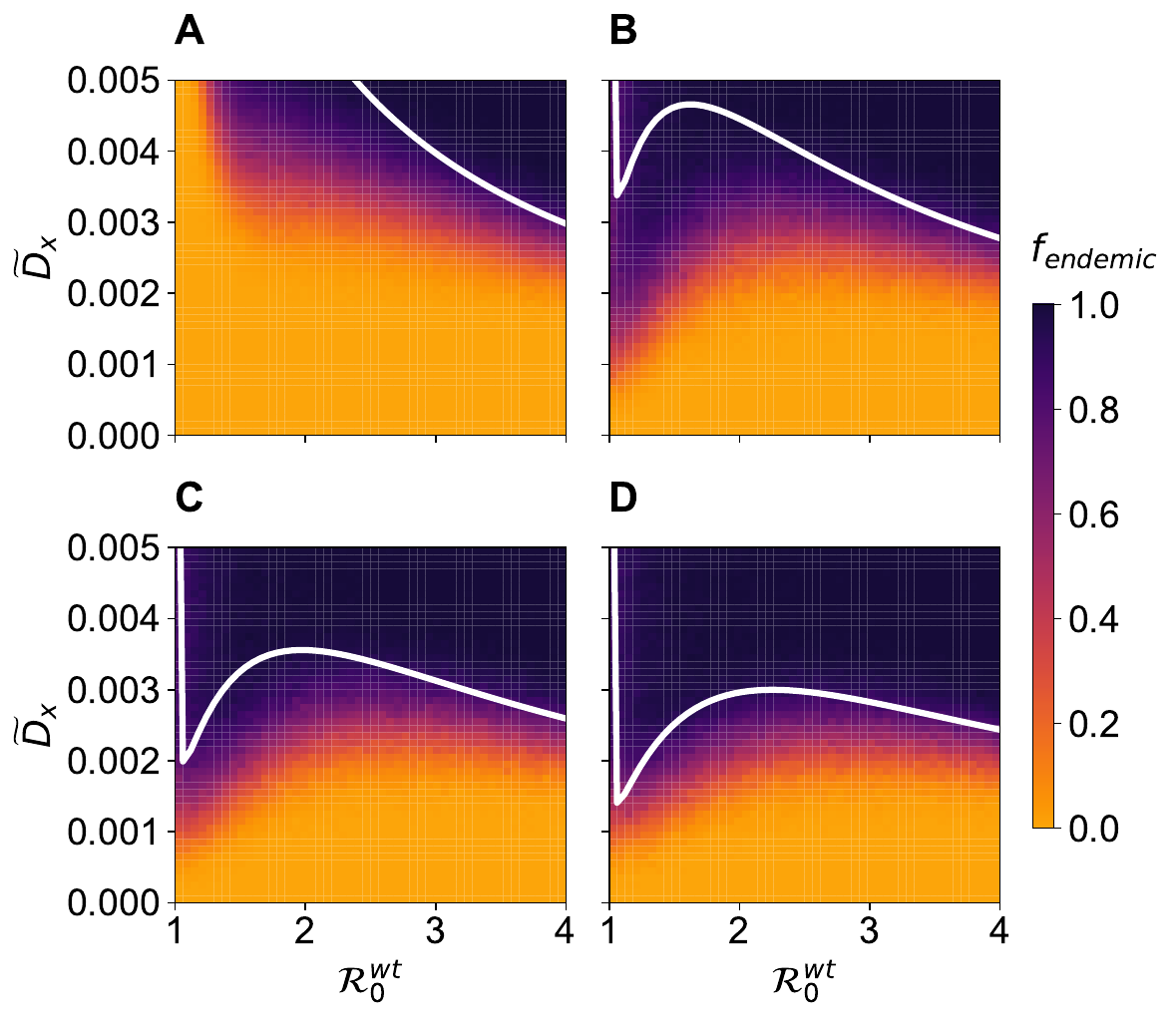}
\caption{{\bf Endemicity of viruses under deterministic evolution.} A)-D): Fraction of endemic realizations $f_{endemic}$ as a function of the basic reproduction number of the wild-type variant $\mathcal{R}^{wt}_0$ and the speed of evolution in the antigenic space $\widetilde{D}_x$. The white solid line shows the theoretical estimation of the critical immunity escape value $\widetilde{D}^C_x$ delimiting the region $f_{endemic}=1$. Such quantity is obtained by setting $a=1/12$ and $b=5$ in Eq.~\ref{eq:Dimmtrans}. The values considered for the speed of evolution in infectiousness are: (A) $\widetilde{D}_\lambda=0$, (B) $\widetilde{D}_\lambda=4\cdot 10^{-5}$, (C) $\widetilde{D}_\lambda=8\cdot 10^{-5}$ and (D) $\widetilde{D}_\lambda=1.2\cdot 10^{-4}$. In all panels, the fraction of endemic realizations is obtained by performing 500 epidemic outbreaks and computing those persisting in the population after $t=1000$ days. The rest of model parameters are the same as in Fig.~\ref{fig:2}.}
\label{fig:4}
\end{figure}
The constant evolution of viral traits eases the mathematical formulation of the model, as all the infected individuals at a given time step $t$ share the same strain $c$, characterized by the antigenic position $x_c(t)=\widetilde{D}_x  t$, where $\widetilde{D}_x$ refers to the (constant) speed of antigenic evolution. Consequently,  the spatiotemporal distribution of infected population across the antigenic space reads $\rho_I(x,t)=\rho_I(t)\delta\left(x-x_c(t)\right)$. Using this expression and considering a travelling wave approach, we can rewrite Eq.~(\ref{eq:det1}) as:
\begin{equation}
    \frac{d\rho_R(\tau)}{d\tau}= - \lambda k \rho_R(\tau)\rho_I(0)\left(1-e^{-\widetilde{D}_x\tau}\right)\ ,
    \label{eq:det2}
\end{equation}
where $\tau = (x_c -x)/\widetilde{D}_x$ refers to the time elapsed since agents' recovery. To estimate the eco-evolutionary constraints shaping the endemicity of virus, we focus on the prevalence of the disease at the epidemic bottleneck after the first epidemic wave, denoted by $\rho_I^{bottleneck}$. When the latter prevalence is small enough, stochastic fluctuations in recovery events drive the system to the absorbing state and the virus becomes extinct, despite its potential to generate another outbreak. Assuming that $\rho_I^{bottleneck}$ remains constant in the bottleneck in a critical scenario, we can derive its value from Eq.~(\ref{eq:det2}) as detailed in Appendix \hyperref[sec:appendixE]{E}, obtaining:
\begin{equation}
    \rho_I^{bottleneck} \simeq \frac{2 \mathcal{R}^{wt}_0 \widetilde{D}_x}{\mu\pi}\ .
\end{equation}
Considering a critical $\rho_I^{bottleneck,C}$ value preventing the stochastic fadeout of simulations, we obtain that the minimum speed of evolution in the antigenic space that a virus requires to become endemic, denoted by $\widetilde{D}_{x}^C$, fulfills:
\begin{equation}
    \widetilde{D}_{x}^C = \frac{\mu a(N)}{\mathcal{R}^{wt}_0}\ ,
    \label{eq:Dc_wotrans}
\end{equation}
where we assume $a(N)=\pi \rho_I^{bottleneck,C}/2$ to depend on the population size $N$, as smaller populations are, in principle, more vulnerable to finite size effects. 

In what follows, we will consider $a$ as a scaling factor to be fitted. To show the validity of this approach, we represent in Fig.~\ref{fig:4}A the fraction of realizations becoming endemic as a function of both $\widetilde{D}_x$ and $\mathcal{R}^{wt}_0$, proving that Eq.~(\ref{eq:Dc_wotrans}) captures the eco-evolutionary constraints allowing the persistence of viruses. As in the original model, in absence of evolution in infectiousness, endemicity monotonically increases with $\mathcal{R}^{wt}_0$. Eq.~(\ref{eq:Dc_wotrans}) also captures the critical surfaces for pathogens with different $\mu$ values (Appendix \hyperref[sec:appendixF]{F}), showing that viruses with shorter infectious windows should evolve more quickly in their antigenic space to achieve endemicity. Likewise, we find that both $\widetilde{D}_x^C$ and, henceforth, $a$ are decreasing functions of the population size $N$, which is an expected result as stochastic fluctuations are less relevant in large populations (Appendix \hyperref[sec:appendixF]{F}).

Now we modify Eq.~(\ref{eq:Dc_wotrans}) to capture how evolution in infectiousness shapes the endemicity of viruses. Under constant evolution, the infectiousness of the circulating strain $c$ at time $t$ reads $\lambda_c (t) = \lambda ^{wt} + \widetilde{D}_\lambda t$. At the epidemic bottleneck, the infectiousness of the virus is determined by the increments accumulated from evolution during the first epidemic wave. Unfortunately, given the non-linearity of the SIR equations, the duration of this phase, denoted by $\tau_{wave}$, cannot be obtained analytically. However, heuristically, we can argue that the characteristic time scale of an epidemic wave is proportional to the time to reach the peak of infected individuals, $\tau_{peak}$, which has been recently estimated~\cite{turkyilmazoglu2021explicit}. Taking into account the evolution in the reproduction number during the first epidemic wave, Eq.~\ref{eq:Dc_wotrans} turns into:
\begin{equation}
    \widetilde{D}_{x}^C = \frac{\mu a(N)}{\mathcal{R}^{wt}_0 + b D_\lambda t_{peak}(\mathcal{R}^{wt}_0,\mu) k/\mu}\ ,
    \label{eq:Dimmtrans}
\end{equation}
where the expression for $t_{peak}$ is provided in Appendix \hyperref[sec:appendixG]{G} and $b$ is another scaling factor to be fitted to obtain the characteristic time scale of the disease, i.e. $\tau_{wave}=b\tau_{peak}$.

Figs.~\ref{fig:4}B-D show how the latter equation fairly captures the non-monotonic behavior of the endemicity with ${\mathcal{R}^{wt}_0}$ for different values of the speed of evolution in infectiousness $D_\lambda$. Note that there are some deviations for $\mathcal{R}_0\simeq 1$, as the expression for $t_{peak}$ neglects the underlying evolution of the virus until reaching the epidemic peak, which is especially relevant in that region of the parameters space. 

More importantly, Eq.~\ref{eq:Dimmtrans} explains why viruses with intermediate infectiousness are less likely to become endemic. For these viruses, ${\cal R}_0^{wt} <  {\mathcal{R}_0^{wt,C}(D_x)}$ but also ${\cal R}_0^{wt} +  b D_\lambda t_{peak}(\mathcal{R}^{wt}_0,\mu) k/\mu<  {\mathcal{R}_0^{wt,C}(D_x)}$, given the short duration of their first epidemic waves. Conversely, the nonlinear inverse dependence of $\tau_{peak}$ on $\mathcal{R}^{wt}_0$ (Appendix \hyperref[sec:appendixG]{G}) makes the duration of the first epidemic wave much longer for pathogens with lower initial infectiousness, thus allowing them to evolve to reach the critical infectiousness before becoming extinct. In Appendix~\hyperref[sec:appendixH]{F}, we analyze in depth the epidemic trajectories obtained from the simulations, finding that viruses with intermediate infectiousness present stronger epidemic bottlenecks and are not able to accumulate neither enough infectiousness nor enough antigenic diversity to generate secondary outbreaks consistently.

\begin{figure}[t!]
\centering
\includegraphics[width=1\columnwidth]{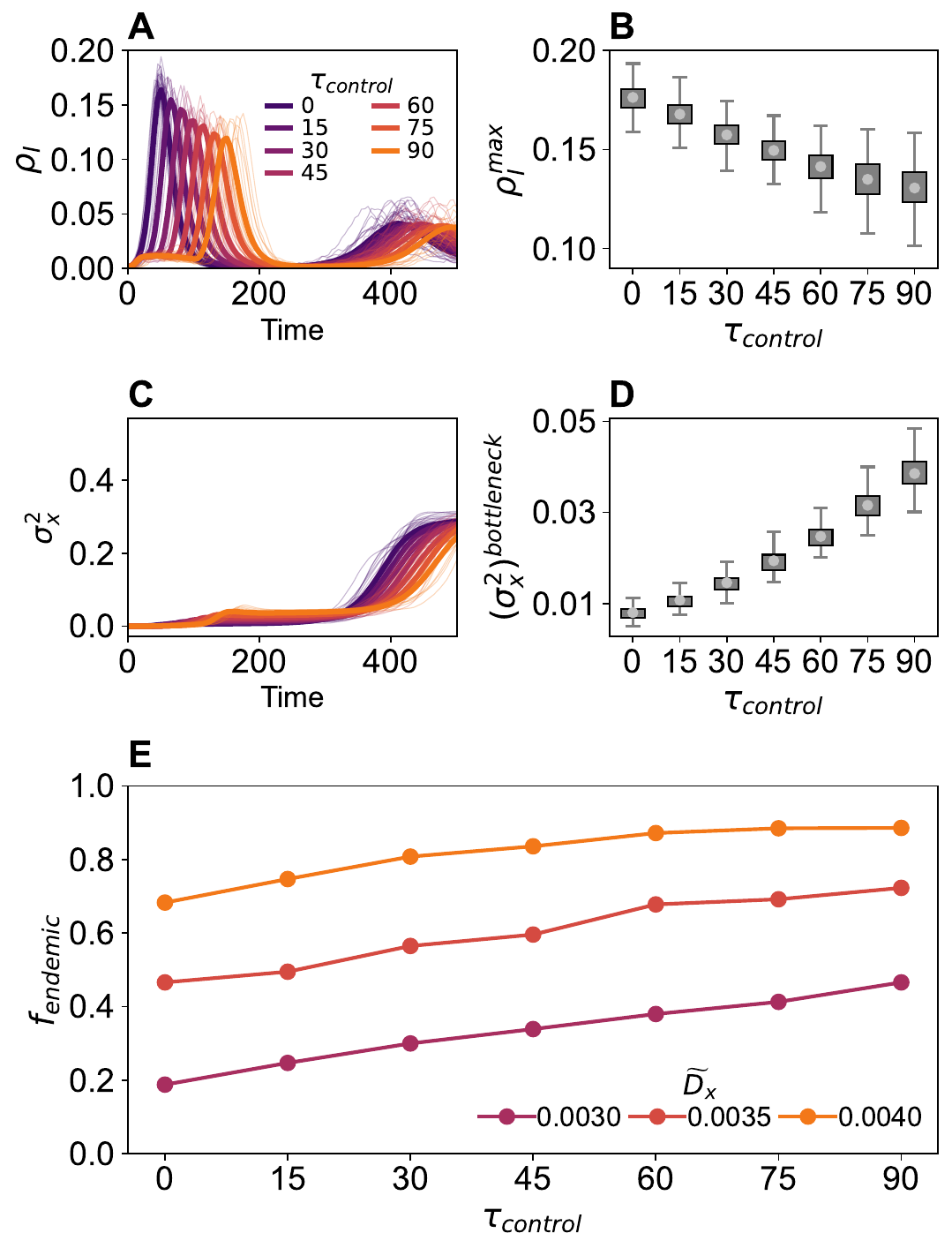}
\caption{{\bf Trade-off between short-term and long-term benefits of control policies}. A): Time evolution of the epidemic prevalence $\rho_I$. Line color denotes the duration of the control policies $\tau_{control}$. B): Distribution of epidemic peak as a function of $\tau_{control}$. C): Time evolution of the antigenic diversity $\sigma_x^2$. Line color denotes the value of $\tau_{control}$. D): Distribution of antigenic diversity at the epidemic bottleneck across endemic realizations, $(\sigma^2_x)^{bottleneck}$, as a function of $\tau_{control}$. We define the epidemic bottleneck as the point with minimum incidence following the first epidemic wave. In panels A) and C) thick lines represent average values across endemic realizations whereas thin lines correspond to single realizations. In panels B) and D), the dot shows the mean of the distribution whereas the whiskers denote its IQR.  In all these panels, the (constant) speed of antigenic evolution is set to $\widetilde{D}_x = 0.0003$. E): Fraction of endemic realizations as a function of $\tau_{control}$. Line color encodes the value of $\widetilde{D}_x$. The criterion used to classify a realization as endemic is the same as the one used in Fig.~\ref{fig:4}. In all panels, we do not consider evolution in infectiousness, i.e. $\widetilde{D}_\lambda=0$ and the epidemiological parameters are the same as in Fig.~\ref{fig:2}.}
\label{fig:5}
\end{figure}

\subsection*{Control policies in the first epidemic wave affects the long-term behavior of epidemics}
Our results reveal that both the evolved infectiousness and the antigenic diversity generated  during the first epidemic wave are crucial to understand viral endemicity. To single out the relevance of antigenic diversity, we now consider viruses not evolving their infectiousness, i.e. $\widetilde{D}_\lambda=0$ and tune the duration of the first epidemic wave by introducing control measures flattening epidemic curves. These control policies are activated when a fraction $\theta = 0.01$ of the population is infected and lifted after $\tau_{control}$ time steps. Among the different choices, we consider that control policies halve the number of contacts of the population, passing from $k=10$ to $k=5$ contacts. Consequently, the reproduction number of the disease is halved with respect to the uncontrolled scenario during the controlled outbreak.

Fig.\ref{fig:5}A shows the infection curves for a virus with ${\cal R}_0^{wt}=2$ and $\widetilde{D}_x=0.003$. As expected, halving the reproduction number renders short-term benefits, for the epidemic peak is reduced as the duration of the policies is increased (see Fig.\ref{fig:5}B). Nonetheless, the effects of such policies on the antigenic diversity $\sigma_x^2$ are more complex (Fig.\ref{fig:5}C). In particular, Fig.\ref{fig:5}D reveals that increasing the duration of control policies gives rise to more antigenic diversity at the epidemic bottleneck. This greater antigenic diversity enables more reinfection events and increases the fraction of endemic realizations (Fig.\ref{fig:5}E), revealing that flattening epidemic curves might enhance the long-term persistence of viruses.

\section*{Discussion}
Understanding the evolutionary forces governing how infectious diseases transition to become endemic in real populations represents a timely research question~\cite{biancolella2022covid}. Here we propose a minimal eco-evolutionary framework to characterize how the evolution in both antigenic and non-antigenic traits shape the course and fate of epidemic outbreaks. Our results from stochastic simulations have first shown that evolution is not constant in both dimensions across time. Instead, the interplay between ecological and epidemiological processes accelerates the evolution in infectiousness at early stages of the outbreak while favoring antigenic diversification once the first epidemic wave is over. This phenomenon occurs because the existence of immune pressure in the population might render mutations reducing infectiousness but changing antigenic properties of viruses beneficial, thus slowing down the overall increase of viral infectiousness. Note that this result cannot be found in Susceptible-Infected-Recovered-Susceptible models with constant waning immunity rates. Empirically, the accelerated evolution in fitness at early generations of RNA viruses has already been reported in culture cells~\cite{novella1995exponential} and characterized theoretically through Fokker-Planck equations~\cite{tsimring1996rna,venegas2014speed}.  Likewise, genomic surveillance data of the SARS-CoV-2~\cite{meijers2023population} reveals the shift from intrinsic selection to antigenic diversification during the COVID-19 pandemic. Yet both evolutionary pathways are in principle uncorrelated, our simulations show that immune pressure in the population also acts a selection force intertwining them, giving rise to universal long-term evolutionary trajectories.

Integrating the evolution of both antigenic and non-antigenic traits drastically changes the eco-evolutionary constraints shaping the endemicity of rapidly evolving viruses. Namely, we have observed how immune escape alone predicts that more infectious pathogens are more likely to become endemic. Conversely, in presence of evolution in infectiousness, viruses with low basic reproduction number, i.e ${\mathcal{R}^{wt}_0}\simeq 1$, are more likely to persist in the population than those with intermediate infectiousness, e.g. ${\mathcal{R}^{wt}_0}\simeq 2$. To characterize analytically this behavior, we have developed a simplistic model assuming a constant evolution in viral traits. Our analysis reveals that the non-monotonic behavior emerges because the latter viruses generate much shorter epidemic outbreaks, thus not being able to accumulate neither enough viral diversity nor enough infectiousness to persist in the population.

The non-monotonic behavior of viral endemicity with respect to the basic reproduction number has already been reported for a SIR model including birth and death dynamics~\cite{parsons2024probability,van1997stochastic}. In this model, the replenishment of the depleted pool of susceptible individuals occurs due to the replacement of dead population by newborn susceptible ones. Interestingly, the non-monotonic behavior emerges theoretically when the duration of the first epidemic wave becomes comparable with the lifetime of individuals. As argued in~\cite{parsons2024probability}, this non-monotonic behavior driven by birth and death dynamics cannot be observed for real diseases, as their typical time scales much shorter than the underlying population dynamics. Sharing the same physical principles, our study provides a more plausible origin of this phenomenon rooted in the replenishment of susceptible population due to antigenic diversification and evolution in infectiousness during the first epidemic wave. Further empirical work will be needed to measure the evolutionary rates of viruses and check whether this non-monotonic behavior is a mere theoretical result or instead can play an important role in real outbreaks.

From a practical point of view, our findings represent a cautionary tale for the implementation of control strategies to manage epidemic scenarios. While the short-term beneficial effects of non-pharmaceutical interventions~\cite{perra2021non,haug2020ranking} cannot be contested, determining their long-term impact on the modified epidemic trajectories still represents a more complex and intricate problem with growing attention in the scientific community~\cite{ashby2023non,gurevich2022modeling,baker2020impact}. Along this line, our results show that controlling an epidemic scenario by reducing the local effective reproduction number ${\cal R}$ might render unforeseen long-term consequences for disease control. In particular, our results show that mitigating an outbreak by flattening the epidemic curves might induce a stronger selection pressure on the virus and increase the risk for their persistence in the population, due to the extended duration of the epidemic outbreak. Likewise, other works have reported that vaccination of the population also alters the antigenic space harboring the different variants of the virus~\cite{aguilar2023impact,miller2021antigenic}, thus potentially shaping the evolutionary strategies observed at the population level~\cite{rouzine2023evolutionary}. Moving from local to global interventions, the evolutionary fingerprint of travel bans and the subsequent geographical genetic isolation of different areas~\cite{wright1943isolation} remains to be solved; phylogeographic analyses~\cite{attwood2022phylogenetic} combined with multiscale epidemic frameworks~\cite{hess1996disease} will be needed to address this question in future works. 

Our findings should be considered in light of several limitations of the minimal framework here introduced. To save computation time, we have implemented discrete-time simulations with synchronous updates rather than using Gillespie-like algorithms. Biologically, the gradual evolution of the virus is challenged by the reported anomalous diffusion behaviors of viruses across the genomic space~\cite{goiriz2023variant} or the complex genotype-phenotype networks shaping the evolution of viruses~\cite{manrubia2021genotypes}. The evolutionary dynamics in our model could also be refined, as we consider that evolution in infectiousness and antigenic position is unbounded and our analytical results rely on a constant evolution in viral traits. To bound the values of the epidemiological parameters, one can explore the effect of different trade-off mechanisms~\cite{goldhill2014evolution}, coupling their joint evolution. Despite the scarce empirical evidence supporting them~\cite{acevedo2019virulence}, evolutionary trade-offs are typically included in models to warrant the existence of evolutionarily stable strategies~\cite{alizon2009virulence}. Our first exploration with the recovery-transmissibility trade-off shows an uneven impact of such mechanism, keeping our findings for some viruses but altering them for others. At the population level, assuming all individuals to be equivalent does not allow our model to capture the strong influence reported for superspreaders~\cite{bendall2023rapid,gomez2021superspreading} or immuno-compromised patients~\cite{sasaki2022antigenic,harari2022drivers} on the eco-evolutionary dynamics of viruses. Moreover, the well-mixed assumption can be improved to include spatially distributed population, as host mobility has been reported to strongly shape the antigenic escape mechanism~\cite{blot2025host}.  Notwithstanding all these limitations, the generality of our findings, based on the relationship between epidemiological and evolutionary time scales, makes us feel confident about their ubiquity in more biologically-grounded models.

In a nutshell, our results underscore the relevance of the interplay between different evolutionary pathways of rapidly evolving viruses in shaping their associated epidemic trajectories. Improving our knowledge on this topic will enhance our preparedness against future epidemic threats and the use of eco-evolutionary frameworks as reliable benchmarks to predict the evolution of emergent pathogens and design optimal control policies to mitigate both their short-term and long-term impact on society. 

\section*{APPENDIX A: Stochastic simulations}
\label{sec:appendixA}
To obtain the epidemic trajectories shown in the manuscript, we perform synchronous discrete-time stochastic simulations using the extended version of the SIR model described in Fig.~\ref{fig:1}, assuming $\Delta t=1$. We split each time step into two different stages governing epidemiological processes and virus evolution respectively. 

At the first stage, we assume that infected individuals recover with a probability  $Pr(I\rightarrow R)= 1-e^{-\mu\Delta t}$. To simulate contagion processes, we assume that each susceptible agent $i$ chooses $k$ neighbors at random and compute the probability of infection as $Pr(S_i\rightarrow I)= 1-e^{-\sum\limits_{l\in \Gamma(i)}\lambda_l\Delta t}$, where $\Gamma(i)$ denotes the set of infected neighbors of agent $i$ at that time step. If a contagion event is successful, the variant contracted by the agent is chosen proportionally to the infectiousness of the variants existing in $\Gamma(i)$. Therefore, the probability that agent $i$ contracts the disease from an infected neighboring agent $j$ is $Pr(S_i\rightarrow I_j)= \dfrac{\lambda_j}{\sum\limits_{l\in \Gamma(i)} \lambda_l} Pr(S_i\rightarrow I)$. For reinfection processes, the same rationale to choose the variant applies, but now considering that the individual rates at which a recovered agent $i$ contracts the disease from an infected agent $j$ is $\lambda^\prime_{ij}=\Theta(x_j-x_i)\lambda_j \left(1-e^{-(x_j-x_i)}\right)$, where $\Theta(x)$ is the Heaviside function introduced to avoid reinfection from past variants. 

Once the epidemiological state of each individual is updated, evolution of the virus only occurs inside infected individuals. Therefore, the infectiousness (antigenic position) of an infected agent $j$ $\lambda_j$ ($x_j$) changes according to $\lambda_j (t+\Delta t)= \lambda_j + D_\lambda\Delta t \mathcal{N} (0,1)$  $\left(x_j (t+\Delta t)= x_j + D_x\Delta t \mathcal{N} (0,1)\right)$, where $\mathcal{N} (0,1)$ represents a random number drawn from the standard normal distribution. Conversely, the antigenic position of recovered individuals remains constant.

\section*{APPENDIX B: Interplay between evolution in infectiousness, immune escape and viral endemicity.}
\label{sec:appendixB}
\begin{figure}[t!]
\includegraphics[width=1\columnwidth]{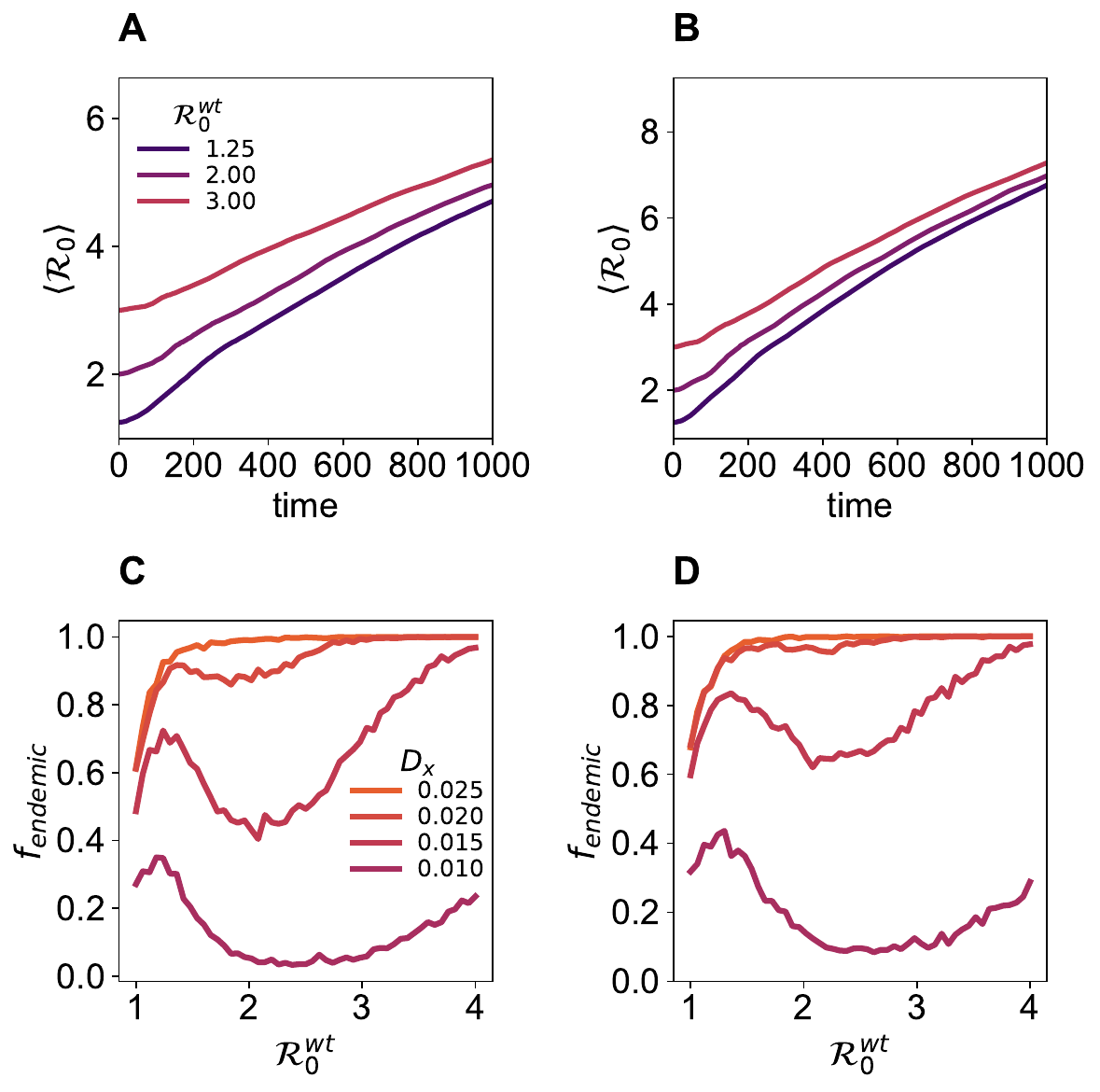}
\caption{{\bf Eco-evolutionary dynamics under stochastic evolution of antigenic and non-antigenic traits.} A)-B): Time evolution of the basic reproduction number ${\cal R}_0$ for endemic epidemic outbreaks. Line color denotes the basic reproduction number of the wild-type variant ${\cal R}_0^{wt}$. The symbol $\langle \cdot \rangle$ denotes that each curve is the result of averaging the individual curves of all endemic realizations observed after simulating $1000$ epidemic outbreaks for each ${\cal R}^{wt}_0$ value. The speed of evolution in the antigenic space is set to $D_x = 0.015$. C-D): Endemicity $f_{endemic}$ of the virus as a function of the basic reproduction number of the wild-type variant ${\cal R}_0^{wt}$. Line color denotes the speed of evolution in the antigenic space $D_x$. In the panels, the values of the speed of evolution in infectiousness used to simulated epidemics are: (A,C) $D_\lambda = 0.0004$ and (B,D) $D_\lambda = 0.0006$. In all the panels, the rest of epidemiological parameters are the same as in Fig.~2 of the main manuscript.}
\label{fig:6}
\end{figure}

In this section, we extend the results reported in Figure~\ref{fig:3} about the influence of the stochastic evolution of both antigenic and non-antigenic traits on epidemic outbreaks.  We start by considering a fixed value of the speed of evolution in antigenic escape ($D_x=0.015$) and different speeds of evolution in infectiousness $D_\lambda$. Figs~\ref{fig:6}A-B confirm that evolution in infectiousness is accelerated at early stages of the outbreak for pathogens with low initial infectiousness. As stated in the main manuscript, this acceleration is more pronounced for high speeds of evolution (Fig~\ref{fig:6}B), giving rise to an earlier collapse of the curves associated to different infectiousness of the wild-type variants. 

We are also interested in determining how evolution in infectiousness shapes the dependence of virus endemicity on the reproduction number of the wild-type variant. Figs.~\ref{fig:6}C-D show that, for $D_{x}=0.015$, the non-monotonic behavior penalizing the endemicity of viruses with intermediate infectiousness appears consistently regardless of the speed of evolution in infectiousness. When antigenic escape gains relevance, i.e. for large $D_x$ values, we instead observe how the non-monotonic behavior disappears as viruses evolve more quickly, yielding the usual increase in endemicity with the infectiousness of the wild-type variant. These results show that changing their intrinsic infectiousness represents an alternative evolutionary pathway for viruses to partially overcome existing immune pressure in the population.

\section*{APPENDIX C: Approximation to the case reproduction number $\mathcal{R}_{case} (t)$} 
\label{sec:appendixC}

In the context of an epidemic outbreak, $\mathcal{R}_{case} (t)$ computes the expected number of contagions made by individuals becoming infected at time $t$. The exact computation of this quantity would require knowing the shape of the distribution of variants in both the infectiousness and antigenic spaces. For the sake of simplicity, we opt for providing a rough estimate of the case reproduction number, $\mathcal{R}^{app}_{case} (t)$. For this purpose, we assume that the standard deviation $\sigma_x$ represents a typical distance observed in the antigenic space and that the mean infectiousness $\bar{\lambda}$ of the circulating variants also constitutes a representative quantity for contagion dynamics. Moreover, let us also neglect the changes in the pool of susceptible and recovered population and the evolution of the virus during the infectious period of the focal agent. Under these assumptions, the approximated case reproduction number can be readily computed as:
\begin{equation}
\mathcal{R}^{app}_{case} (t) = \frac{\bar{\lambda}}{\mu}\left[\rho_S(t) + \rho_R(t) \left(1-e^{-\sigma_x}\right)\right]\ ,
\end{equation}
where $\rho_m(t)$ represents the fraction of population in state $m$ at time $t$. 

\section*{APPENDIX D: Transmission-recovery tradeoff}
\label{sec:appendixD}

The model presented in the main manuscript does not include any biological trade-off linking the different epidemiological parameters of pathogens and, therefore, their evolution. These trade-offs are widely adopted in the modelling community to produce evolutionarily stable strategies (ESS) bounding the evolution of viruses. Among them, the virulence-transmissibility trade-off is undoubtely the most studied one following the seminal work by Anderson \& May~\cite{anderson1982coevolution}. The virulence-transmissibility trade-off assumes that the virulence of a virus, defined as the rate at which the host die because of the pathogen, increases with the infectiousness of the virus. This increased lethality shortens the infectious period during which the virus host can transmit the pathogen to other individuals. This double-edge sword creates an optimal strategy at intermediate infectiousness, maximizing the expected size of the offspring produced by infected individuals in next generations. 

As our framework does not incorporate any birth-death processes involving hosts, we opt for not including the virulence-transmissibility trade-off. Instead, we can study how the joint evolution in infectiousness and in the infectious period affects the eco-evolutionary dynamics of pathogens by introducing the so-called transmission-recovery trade-off~\cite{alizon2008transmission}. This trade-off implies that the infectious period is shortened for highly transmissible pathogens and emerges naturally when both the production of immune defenses against a pathogen and its inter-host infectiousness are assumed to be proportional to the pathogen load inside hosts. 

\begin{figure}[t!]
\centering
\includegraphics[width=1\columnwidth]{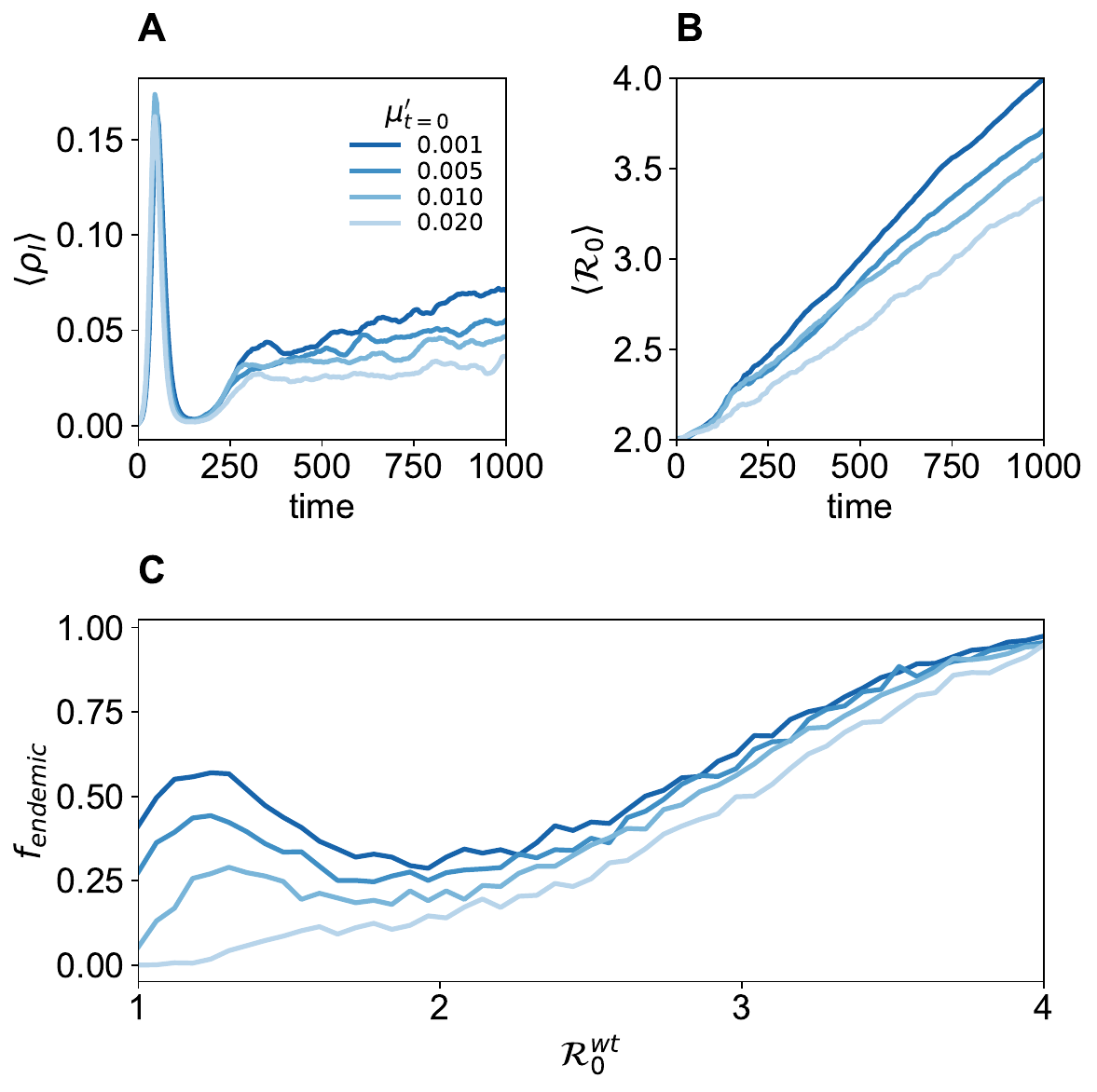}
\caption{{\bf Impact of the transmission-recovery trade-off on eco-evolutionary dynamics}. A)-B): Time evolution of different epidemiological quantities and virus traits. The quantities shown correspond to: (A) fraction of infected population $\rho_I$ and (B) basic reproduction number ${\cal R}_0$. The symbol $\langle \cdot \rangle$ denotes that each curve is the result of averaging the individual curves of all endemic realizations observed after simulating $1000$ epidemic outbreaks for each ${\cal R}_0$ value. The speeds of evolution in the infectiousness and antigenic spaces are set to $D_\lambda=0.0003$ and $D_x=0.015$ respectively. C): Endemicity $f_{endemic}$ of the virus as a function of the basic reproduction number of the wild-type variant ${\cal R}_0^{wt}$. In all the panels, line color denotes the initial value of the evolving recovery rate $\mu^\prime_{t=0}$ and the rest of epidemiological parameters are the same as in Fig.~\ref{fig:2}.}
\label{fig:7}
\end{figure}

To account for the transmission-recovery trade-off, we decompose the recovery rate of individuals $\mu$ into two components: the baseline recovery rate $\mu_0$, constant and independent of the pathogen load, and the recovery rate $\mu^\prime$ changing over time as a result of the evolution in infectiousness $\lambda$ and the aforementioned trade-off. Mathematically, we assume $\lambda=\gamma\sqrt{\mu^\prime}$, where $\gamma$ is a constant factor related to the basic reproduction number of the wild-type variant ${\cal R}^{wt}_0$ and the initial value of the evolving recovery rate $\mu^\prime_{(t=0)}$. Namely:
\begin{equation}
    \gamma = \frac{{\cal R}_0^{wt}\left(\mu_0 + \mu^\prime_{(t=0)}\right)}{k \sqrt{\mu^\prime_{(t=0)}}}\ .
\end{equation}
Let us first consider viruses with ${\cal R}_0^{wt}=2$ evolving both their infectiousness ($D_\lambda=0.0003$) and their antigenic position ($D_x = 0.015$). To analyze the impact of the transmission-recovery trade-off, we fix the baseline recovery rate to $\mu_0=1.0/7$ days$^{-1}$ and study the impact of varying the initial value of the evolving recovery rate, $\mu^\prime_{(t=0)}$, on the epidemic curves observed in the population. When the transmission-recovery trade-off plays a minor role, i.e. when $\mu^\prime_{(t=0)} \ll \mu_0$, the epidemic curves represented in Fig.~\ref{fig:7}A resemble those shown in Fig.~\ref{fig:3}, characterized by a first prominent epidemic wave followed by an epidemic bottleneck and an endemic regime with a steady growth rate in the number of cases. However, as both rates become comparable, the trade-off bounds the evolution of the virus and considerably slows down the growth rate of cases in the endemic regime, yielding a roughly constant prevalence of the disease for some epidemic scenarios. This deceleration also affects the evolution of the basic reproduction number ${\mathcal{R}_0}$, which is slower as the trade-off gains more relevance (Fig.~\ref{fig:6}B). 

The transmission-recovery trade-off alters the epidemiological time scales of evolving viruses by shortening the infectious period of infected individuals. This phenomenon also changes the features of the viruses more prone to persist in populations (Fig.~\ref{fig:7}C). Namely, the non-mononotic behavior of viral endemicity with ${\cal R}_0^{wt}$ is preserved when $\mu^\prime_{(t=0)} \ll \mu_0$. Conversely, when $\mu^\prime_{(t=0)} \simeq \mu_0$, viral endemicity always increases with the basic reproduction number of the wild-type variant. We believe this phenomenon arises from the shortening of the infectious period and the slowing down in the evolution of infectiousness, making it difficult for weakly infectious pathogens to reach the critical values of the basic reproduction number needed to become endemic. 

\section*{Appendix E: Epidemic prevalence at equilibrium with constant antigenic escape}
\label{sec:appendixE}
As shown in the main manuscript, assuming a constant evolution of the antigenic position allows getting an ODE, Eq.~(\ref{eq:det2}) governing the time evolution of the density of recovered population $\rho_R (\tau)$, where $\tau$ refers to the time elapsed since the recovery of an agent. Considering a steady prevalence of the disease at the epidemic bottleneck, i.e. $\rho_I(0)=\rho_I^{bottleneck}$, and a regime of strong immune pressure, i.e. $\widetilde{D}_x\tau\ll 1$, we solve the previous ODE obtaining:
\begin{equation}
 \rho_R(\tau) = \mu \rho_I^{bottleneck} e^{-\widetilde{D}_x\lambda k \rho_I^{bottleneck} \tau^2/2}\ ,
\end{equation}
where we have used $\lim_{\tau \to 0}\rho_R(\tau)=\mu \rho_I^{bottleneck}$. To find the epidemic prevalence, we should recall that no susceptible individuals remain in the endemic equilibrium and use the normalization condition $\int \rho_R(\tau)d\tau + \rho_I^{bottleneck} = 1$, yielding:
\begin{equation}
    \sqrt{\frac{\mu\pi \rho_I^{bottleneck}}{2\mathcal{R}_0 \widetilde{D}_x}} + \rho_I^{bottleneck}=1 \ .
\end{equation}
Assuming a very small prevalence in the endemic regime, i.e. $\rho_I^{bottleneck}\ll 1$, we can estimate $\rho_I^{bottleneck}$ as
\begin{equation}
    \rho_I^{bottleneck} \simeq \frac{2 \mathcal{R}_0 \widetilde{D}_x}{\mu\pi}\ .
\end{equation}

\section*{Appendix F: Influence of infectious period and population size on $\widetilde{D}^C_x$}
\label{sec:appendixF}
\begin{figure}
\centering
\includegraphics[width=\columnwidth]{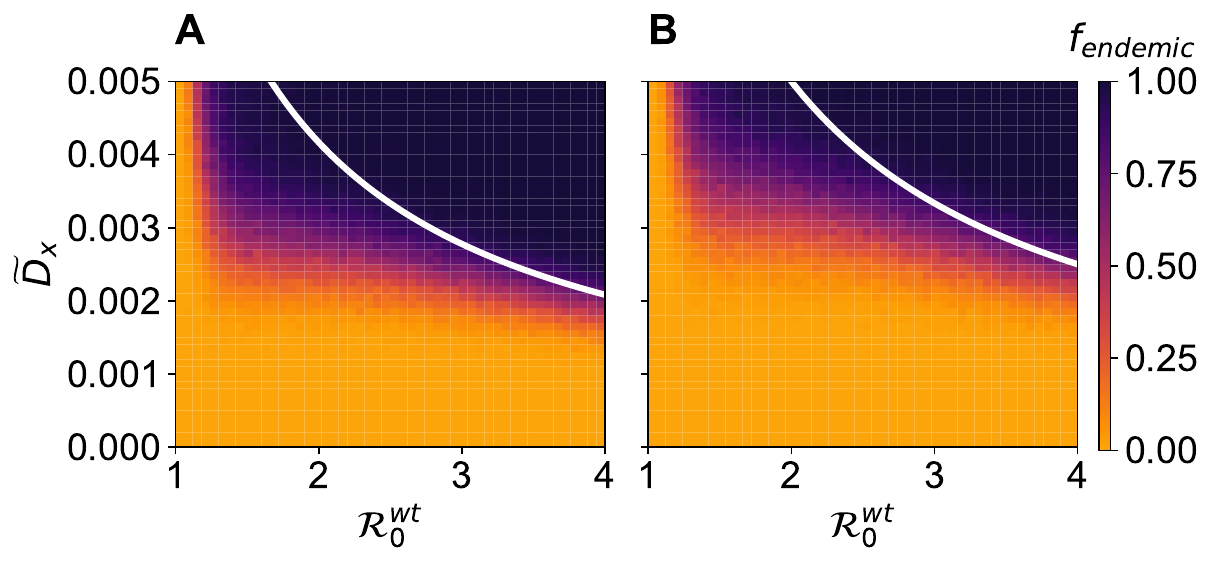}
\caption{{\bf Endemicity of viruses under deterministic evolution.} A)-B): Fraction of endemic realizations $f_{endemic}$ as a function of the basic reproduction number of the wild-type variant $\mathcal{R}^{wt}_0$ and the speed of evolution in the antigenic space $\widetilde{D}_x$. The white solid line shows the theoretical estimation of the critical immunity escape value $\widetilde{D}^C_x$ delimiting the region $f_{endemic}=1$. Such quantity is obtained by setting $a=1/12$ and $b=5$ in Eq.~4 of the main text. The values considered for the duration of the infectious windows are: (A) $\mu^{-1}=10$ and (B) $\mu^{-1}=8$ days. In all panels, the speed of evolution in infectiousness is set to $\widetilde{D}_\lambda=0$. Moreover, the fraction of endemic realizations by performing 500 epidemic outbreaks and computing those persisting in the population after $t=1000$ days. The rest of model parameters are the same as in Fig.~2 of the main manuscript.}
\label{fig:8}
\end{figure}

Eq.~(\ref{eq:Dc_wotrans}) represents the critical evolution in the antigenic position $\widetilde{D}^C_x$ for the endemicity of viruses not evolving their infectiousness . While in the main manuscript we focus on showing the inverse dependence of $\widetilde{D}^C_{x}$ on $\mathcal{R}^{wt}_0$, here we are interested in exploring how the boundary of the endemic regime depends on the rest of parameters. To address this question, we first set the population size to $N=10000$, consequently fixing the constant $a=1/12$ (Fig.~\ref{fig:4}A), and represent in Fig.~\ref{fig:8} the fraction of endemic realizations as a function of the reproduction number ${\cal R}_0^{wt}$ and the immune escape $\widetilde{D}_{x}$ for two additional values of the duration of the infectious window $\mu^{-1}$:  $\mu^{-1}=8$ days (Fig.~\ref{fig:8}A) and $\mu^{-1}=10$ days (Fig.~\ref{fig:8}B). There we check that the theoretical estimation provided by Eq.~\ref{eq:Dc_wotrans} fairly captures the endemic regime for the three epidemic scenarios and confirm that the critical value $\widetilde{D}^C_{x}$ increases as the infectious period is reduced.

\begin{figure}
\centering
\includegraphics[width=\columnwidth]{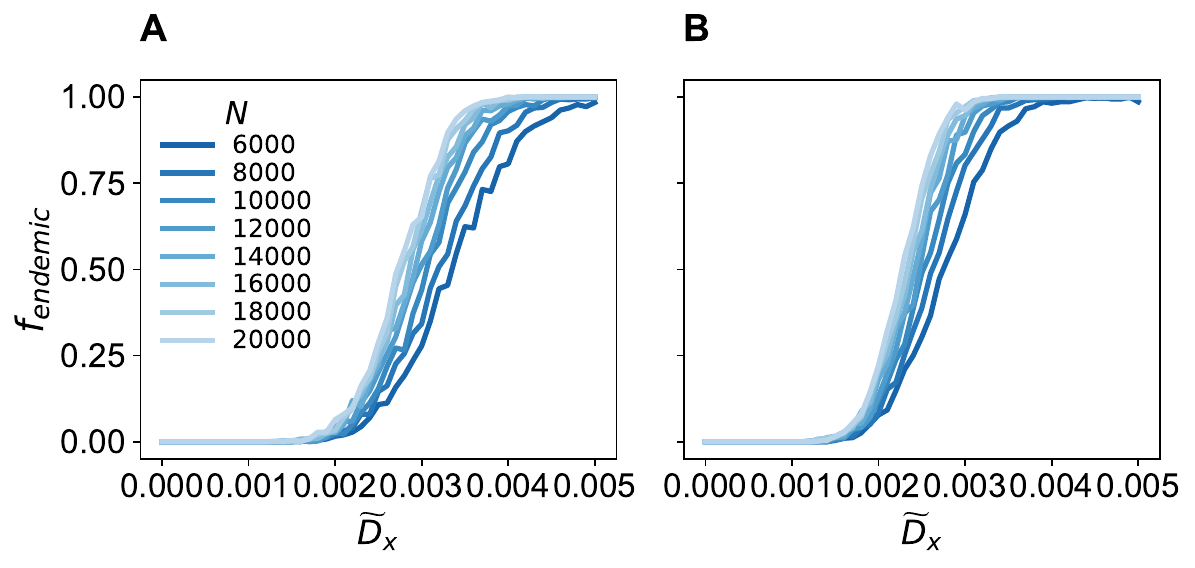}
\caption{{\bf Finite size effects on the endemicity of viruses under deterministic evolution.} Endemicity $f_{endemic}$ of the virus as a function of the basic reproduction number of the speed of evolution in the antigenic space $\widetilde{D}_x$. Two different values for the reproduction number of the wild-type variant are considered: (A) $\mathcal{R}_0^{wt}=3$ and (B) $\mathcal{R}_0^{wt}=4$. In all panels, line color denotes the population size $N$ and the speed of evolution in infectiousness is set to $\widetilde{D}_\lambda=0$ and the rest of epidemiological parameters are the same as in Fig.~2 of the main manuscript.}
\label{fig:9}
\end{figure}

To estimate finite size effects on the virus endemicity, we represent in Figure~\ref{fig:9} how the fraction of endemic realizations changes a function of the immune escape $\widetilde{D}_{x}$ and the population size $N$ for two pathogens differing in their reproduction number: $\mathcal{R}_0^{wt}=3$ (Fig.~\ref{fig:9}A) and 
$\mathcal{R}_0^{wt}=4$ (Fig.~\ref{fig:9}B). In both panels, we clearly show how the critical immune escape for endemicity decreases as populations become larger. Considering Eq.~(\ref{eq:Dc_wotrans}), this result reveals an inverse dependence of the scaling factor $a$ with the population size $N$. Namely, for small population sizes, a greater epidemic prevalence at the bottleneck is needed to prevent the extinction of epidemic outbreaks. Interestingly, the curves seem to collapse for large population sizes ($N\simeq 20000$), thus highlighting that $a(N)$ reaches an asymptomatic value for large population. The latter fact highlights that finite size effects are less relevant in large populations, for which transitions between endemic and non-endemic outbreaks are mainly driven by the interplay between epidemiological and evolutionary time scales.

\section*{Appendix G: Time for an epidemic peak in a SIR model}
\label{sec:appendixG}
For a well-mixed population with constant size, the time to the epidemic peak $t_{peak}$ in the standard SIR model can be approximated as~\cite{turkyilmazoglu2021explicit}:
\begin{equation}
    t_{peak} (\mathcal{R}^{wt}_0,\mu,i_0)=\frac{1}{\mu} \frac{1}{\mathcal{R}^{wt}_0-1} \ln \left(\mathcal{R}^{wt}_0+i_0^{-1}\left(\mathcal{R}^{wt}_0-1\right)^2\right)
    \label{eq:tpeak}
\end{equation}

where $i_0$ refers to the fraction of population initially infected. This expression represents a first-order approximation to the exact computation of the epidemic peak in the standard SIR model. Nonetheless, it provides reasonably accurate estimates of the peak in most of the epidemic scenarios~\cite{turkyilmazoglu2021explicit}. In our model, the latter expression is used to obtain a characteristic time scale during which viruses can evolve their transmissibility to avoid getting extinct. Consequently, note that the former expression comes with intrinsic limitations, as it assumes a fixed value of the infectiousness $\lambda$ and not a evolving quantity over time. Therefore, we should expect predictions to be misaligned with the actual epidemic peak for quickly evolving viruses with low initial infectiousness, i.e. ${\mathcal{R}_0\simeq 1}$. In that region, the slow propagation of the virus leads to a substantial evolution in the infectiousness of the virus throughout the first epidemic wave, which in turn produces a much shorter epidemic wave than the one predicted by Eq.~(\ref{eq:tpeak}).

\begin{figure}[t!]
\centering
\includegraphics[width=\columnwidth]{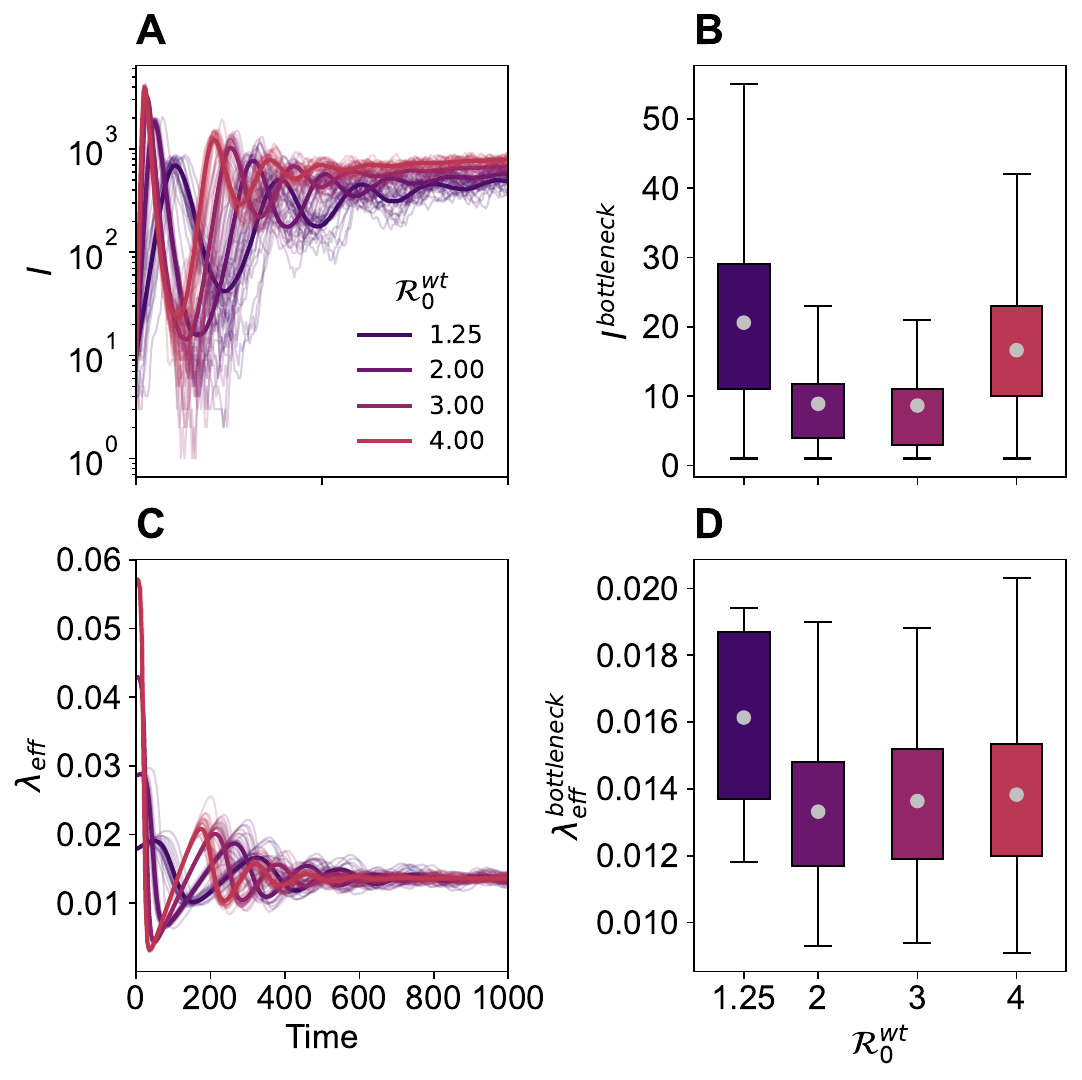}
\caption{{\bf Non-monotonic behavior of epidemiological indicators at the bottleneck with ${\cal R}_0^{wt}$.} A): Time evolution of number of infected individuals $I$ for endemic epidemic outbreaks. B): Distribution of the minimum number of infected individuals at the epidemic bottleneck $I^{bottleneck}$ across endemic realizations as a function of the basic reproduction number of the wild-type variant ${\cal R}_0^{wt}$. C): Time evolution of effective infectiousness $\lambda_{eff}$ for endemic epidemic outbreaks. D): Distribution of the effective infectiousness at the epidemic bottleneck $\lambda^{bottleneck}_{eff}$ across endemic realizations as a function of ${\cal R}_0^{wt}$. In all panels, color encodes ${\cal R}_0^{wt}$. In panels (A) and (C), thick lines show average values whereas single lines correspond to single realizations. in panels (B) and (D), the grey dot denotes the average across endemic realizations whereas the whiskers of the boxplot denote IQR. In all the panels, the speeds of constant evolution have been set to $\widetilde{D}_x = 0.003$ and $\widetilde{D}_\lambda = 0.00004$ respectively. The rest of epidemiological parameters are the same as in Fig.~\ref{fig:2}}
\label{fig:10}
\end{figure}

\section*{Appendix H: Non-monotonic behavior of epidemiological parameters in the bottleneck}
\label{sec:appendixH}

In the main text, we find analytically an estimation of the critical speed of (constant) antigenic evolution $\widetilde{D}_x^C$ for a virus becoming endemic. This mathematical expression captures the non-monotonic behavior of viral endemicity with the infectiousness of the wild-type variant ${\cal R}_0^{wt}$. In this section, we explore whether there is a fingerprint of this non-monotonic behavior in the infection and evolution curves obtained from the simulations. Assuming $\widetilde{D}_x = 0.003$ and $\widetilde{D}_\lambda = 0.00004$, we present in Fig.~\ref{fig:10}A, the epidemic curves obtained for different values of ${\cal R}_0^{wt}$. One salient feature of these curves is that the prevalence at the epidemic bottleneck strongly varies with ${\cal R}_0^{wt}$. We characterize this phenomenon in Fig.~\ref{fig:10}B, where we show that there exists a non-monotonic behavior of the size of infected population with ${\cal R}_0^{wt}$. The latter makes outbreaks generated by viruses with intermediate infectiousness more likely to become extinct due to stochastic fluctuations.

In addition to the epidemic prevalence at the bottleneck, we also define here the effective infectiousness of the virus $\lambda_{eff} (t)$ as the product between the infectiousness of the circulating variant $\lambda_c$ and the fraction of population exposed of the disease at time $t$. Taking into account that all the infected population share the same strain under constant population, this indicator reads:
\begin{equation}
\lambda_{eff} (t)= \sum_{i=1}^N \lambda_c(t) \left[\delta_{\Gamma_i (t),S} + \delta_{\Gamma_i (t),R} \left(1-e^{-(x_c(t) - x_i)}\right)\right]\ ,
\end{equation}
where $\delta$ denotes the Kronecker delta and $\Gamma_{i}(t)$  the state of the agent $i$ at time $t$ during a given outbreak. Fig.~\ref{fig:10}C shows the evolution of $\lambda_{eff}$ for multiple outbreaks generated by viruses differing in ${\cal R}_0^{wt}$. We see that the model with constant evolution retrieves the universal long-term behavior of the evolution curves observed in the original model. Focusing on the epidemic bottleneck, Fig.~\ref{fig:10}D reveals that the effective infectiousness of the virus at the bottleneck also display a non-monotonic behavior with ${\cal R}_0^{wt}$. As stated in the manuscript, this arises from the trade-off between the original infectiousness of the viruses and the increments accumulated in their first epidemic waves. Taken together, our results show how viruses with ${\cal R}_0^{wt}$ are more likely to disappear due to the strong population bottlenecks after the first epidemic wave and an insufficient evolved effective infectiousness to create a new secondary outbreak.

\section*{Code and data availability statements}
No new data have been generated in this article. Code needed to reproduce the results is publicly available at~\cite{githubrepository}.

\begin{acknowledgments}
The author thanks A. Alsina, S. Lamata-Ot\'in, J. G\'omez-Garde\~nes, C. G\'omez-Ambrosi and A. Arenas for their extremely valuable feedback on the results here discussed and their careful reading of the first version of the manuscript. The author also thanks L. Arola, R. Vishwakarma, P. Castioni and L. Fant for insightful discussions about the paper. The author acknowledges financial support through grants JDC2022-048339-I and PID2021-128005NB- C21 funded by MCIN/AEI/10.13039/501100011033 and the European Union “NextGenerationEU”/PRTR”.
\end{acknowledgments}

\clearpage
\newpage
\newpage

\end{document}